\documentclass{article}
\begin{document}

\noindent{\bf NASA tests of Einstein's Universe call for non-empty space physics of nonlocal classical matter}

\bigskip
\noindent
I.E. Bulyzhenkov-Widicker
 
 \noindent{\small {\it Institute of Spectroscopy, Troitsk, Russia and University of Ottawa,  Ottawa,  Canada}}


\bigskip
\noindent {\small {\bf Abstract}. LLR and GPB geodetic precessions have tested flatness for Einsten-Infeld-Hoffmann dynamics of distributed relativistic matter.}  

\bigskip



The non-Newtonian geodetic precession of the asymmetrical Earth-Moon gyroscope in the gravitational fields of the Sun was first noted by de Sitter \cite{Sit} in 1916. The modern quantitative analysis of the lunar-laser-ranging (LLR) data, accumulated by NASA between 1970 and 1986, was finally revealed \cite{Sha} in 1988. These data have been interpreted through the post-Newtonian parameter $\gamma$ related to supposed departure of 3D space from Euclidean flatness. Bending of empty space should unavoidably accompany point source models employed by Schwarzschild and Droste in the 1916 solutions \cite{Schw} to Einstein's 1915 equation. Nowadays everyone agrees that the energy-momentum tensor density, rather than point mass, is a real source of General Relativity (GR) fields in this nonlinear gravitational equation for classical matter. From here Newtonian radial field energy density, $ \propto (\nabla U)^2 $,  works as a nonlocal $r^{-4}$ material source of gra\-vi\-tation. In other words, realistic space-time-energy organizations in Einstein's Universe cannot operate in principle with the empty (or free) space paradigm, because the GR energy-source is distributed along with its field  over  non-empty space filled everywhere by gravitational energy. 

Nonetheless, the numerical coincidence (2\% experimental error in 1988 and 0.7\% in 1996) of measurements with the Schwarzschild parameter $\gamma \equiv 1$ for the pre-quantum point particle was accepted by NASA researchers as an ultimate proof of space curvature that stroked out flat-space proponents from main gravitational conferences. Sommerfeld, Brilluen, Schwinger and many other classics had already explained the conceptual role of Euclidean space for modern physics.  Feyman even insisted on conference registrations under a pseudonym (for example in 1957, Chapel Hill, NC) to express his distain to unphysical interpretation of gravitational phenomena. In 1939 Einstein \cite{Ein} (and Narlikar\cite{Nar} in 1985) finally rejected the unrealistic Schwarzschild's solution, but the black hole generation of cosmologists is persistently ignoring `irrelevant' criticism of their `reluctant father'.

Recent data of the Gravity Probe B mission \cite{Eve} could be also gladly accepted by the Schwarzschild-Schiff model with ${\bf \Omega}_G = (2^{-1}+\gamma){\bf v}\times {\bf \nabla} U $ for the geodetic precession frequency of four-vector spins without radial structures in locally curved space \cite{Sch}.  The superficial modeling of  (distributed) classical bodies through point masses in empty-space  prevents timely recognition of the GPB ground-breaking finding - 3D space is strictly flat (better than 1\%) with respect to translations and rotations. The point is that the first GPB data (einstein.stanford.edu) for small superconducting gyroscopes reiterate the same geodetic precession of the Moon-Earth gyroscope. This distributed system is well discussed without point singularities and its slow-motion precession is related in Einstein's relativity to inhomogeneous time dilatation over the Moon orbit, rather than to local space curvature in question at the gyroscope center of inertia. 

Recall that Weyl submitted his correct computations for non-point relativistic tops in 1923, well before the Einstein-Infeld-Hoffmann equation \cite{EIH}  of slow relativistic motion was derived in 1938. The similar post-Newtonian equations for slowly moving and rotating GR systems having finite dimensions and masses were also obtained by other relativists \cite{Edd}. And the classical Lagrange formalism for the Einstein-Infeld-Hoffmann dynamics very clearly specifies the enhanced geodetic precession of non-point orbiting gyroscopes through GR's time dilatation or the $g_{oo}$ metric component \cite{Lan}. 

Why was the known time-dilatation nature of the de Sitter - Weyl - Einstein - Infeld - Hoffmann geodetic precession never mentioned by NASA researchers as the original GR alternative to their curve-space interpretations? Einstein never refused from his 1938 post-Newtonian dynamics of distributed passive-active elements and tried to introduce non-point gravitational sources even into the 1915 covariant equation.  Initially Einstein and Grossmann put Newtonian potential $U = - GM/r$ only into the time subinterval in the Minkowski space-time interval \cite {Gro}. Lately Schwarzschild's point source constructions reconnected both subintervals with the gravitational potential and badly redirected the Einstein-Grossmann metric project into empty-space frames of pre-quantum physics. 

Again, why were very strict anti-Schwarzschild statements of Einstein never cited by NASA investigators of `Einstein's Universe'?  Unrealistic point sources or spins are very useful sometimes for simplified, model interpretation of physical reality. But they should not substitute or strike out more rigorous pro-Einstein approaches to self-organization of space-time-matter in the nonlinear GR equations with nonlocal energy-charges or with non-empty space \cite{Bul}.   All GR flowers should blossom and all solutions of the Einstein equation for empty and non-empty spaces should be equally discussed and respectfully compared by GRG19 and other scientific forums. NASA reports cannot  ignore Einstein-Infeld-Hoffman physics for slowly-rotating distributions of mass-energy in favor of Schwarzschild-Schiff mathematics for point particle-spins in question. What are the reasons to modify Einstein's physics prior to its tests? 

 In my view, NASA's LLR and GPB public releases have perfectly confirmed GR's time-dilatation in the Einstein-Infeld-Hoffmann approach to distributed GPB gyroscopes without any anti-Einstein contributions from non-existing spatial curvature. There is no need to reinterpret the classical GR rotation through point-spin innovations. I would propose to keep the 1923 Weyl-Einstein non-point gyroscope and to compare the `Entwurf' flat-space generalization of the Minkowski interval \cite{Gro, Bul} with the LLR and GPB data for spin-orbit and spin-spin frame-dragging. Any energy density may be conjugated only with local time, not with space coordinates. Therefore I call for non-empty space interpretation of Einstein's physics for the global Machian overlap of $r^{-4}$ nonlocal  matter in the Euclidean Universe. A web workshop `Nonlocal Classical Matter in General Relativity and Cosmology' of anonymous and open Schwarzschild critics, including condensed matter physicists, would be beneficial for separation of realistic Einstein's heritage from point-mass simplifications and popular empty-space models. 


\end {document}